\documentclass[pdflatex,sn-mathphys-num]{sn-jnl}% Math and Physical Sciences Numbered Reference Style
\usepackage{graphicx}%
\usepackage{multirow}%
\usepackage{amsmath,amssymb,amsfonts}%
\usepackage{amsthm}%
\usepackage{mathrsfs}%
\usepackage[title]{appendix}%
\usepackage{xcolor}%
\usepackage{textcomp}%
\usepackage{manyfoot}%
\usepackage{booktabs}%
\usepackage{algorithm}%
\usepackage{algorithmicx}%
\usepackage{algpseudocode}%
\usepackage{listings}%
\usepackage{caption}
\usepackage{subcaption}

\theoremstyle{thmstyleone}%
\theoremstyle{thmstyletwo}%

\theoremstyle{thmstylethree}%

\begin{document}

\title[Optimizing VQE Ansatz for Studying Tight-Binding Models with \textit{sd}-Interaction and On-Site Coulomb Repulsion]{Optimizing VQE Ansatze for Studying Tight-Binding Models with \textit{sd}-Interaction and On-Site Coulomb Repulsion}

%%=============================================================%%
%% GivenName	-> \fnm{Joergen W.}
%% Particle	-> \spfx{van der} -> surname prefix
%% FamilyName	-> \sur{Ploeg}
%% Suffix	-> \sfx{IV}
%% \author*[1,2]{\fnm{Joergen W.} \spfx{van der} \sur{Ploeg} 
%%  \sfx{IV}}\email{iauthor@gmail.com}
%%=============================================================%%

\author*[1]{\fnm{Oleg} \sur{Udalov}}\email{udalovog@gmail.com}

\affil*[1]{\orgdiv{Independent Researcher}, \orgaddress{\postcode{20878}, \state{MD}, \country{USA}}}

%%==================================%%
%% Sample for unstructured abstract %%
%%==================================%%

\abstract{The VQE algorithm is applied to the problem of finding the ground state of a lattice model with on-site Coulomb repulsion, nearest-neighbor hopping, and on-site sd-interaction. We compare the performance of several ansatze, including cluster and generic forms. Several modifications of the standard cluster ansatz implementation are proposed, which significantly reduce the number of two-qubit gates. Different classical optimizers are employed within the VQE algorithm. The performance of the algorithms is evaluated using both noiseless and noisy simulations.}

%%================================%%
%% Sample for structured abstract %%
%%================================%%


%%\pacs[JEL Classification]{D8, H51}

%%\pacs[MSC Classification]{35A01, 65L10, 65L12, 65L20, 65L70}

\maketitle

\section{Introduction}\label{sec1}

Nowadays, quantum computer (QC) simulations of physical systems attract increasing attention due to the growing availability of quantum hardware. Multiple companies now provide access to their QCs. Although the computational power of these devices is still far below that of classical counterparts, the promise of future advantages motivates research into this novel computational paradigm and the exploration of new quantum algorithms along with their limitations.

One of the most promising algorithms for today's noisy QCs is the variational quantum eigensolver (VQE)~\cite{Tilly2022VQEReview,Peruzzo2014VQE,Yung2014FromDigital,Barends2015Digital,McClean2016Theory,Wang2018Generalized,Cerezo2021Variational}. VQE enables the simulation of physical systems and the determination of both ground and excited states. Its most notable application is in molecular simulations~\cite{Romero2018Strategies,PhysRevA.98.022322,McArdle2018QuantumComputationalChemistry}, while another common use case is in lattice models~\cite{PRXQuantum.1.020319,Stanisic2022Observing,PhysRevResearch.4.023190,PhysRevA.111.052619,Shalini2024FermiHubbard}.

The VQE algorithm consists of several key components: (1) an initial guess for the wave function, (2) an ansatz for constructing the wave function, (3) a classical optimizer, and (4) a measurement block. In this report, we focus on the first three components. We apply VQE to study a tight-binding model with on-site Coulomb repulsion and sd interaction. We compare several different ansatze, optimizers, and initial states to determine the most effective combination for the proposed physical problem. For the ansatz, we investigate different versions of the cluster ansatz~\cite{PhysRevA.92.042303,Lee2018GeneralizedUCCpreprint} and propose modifications aimed at reducing circuit depth. We also benchmark their performance against generic ansatze~\cite{Sim2019Expressibility,BravoPrieto2020Scaling,Kandala2017HardwareEfficientVQE}. Finally, we use both noiseless and noisy simulations to assess algorithmic performance.

The influence of the initial guess on VQE convergence and the fidelity of the obtained wave function was studied in Ref.~\cite{PhysRevB.109.035128}, where it was demonstrated that the choice of initial guess is critically important. Motivated by this, we also employ various initial guesses in our simulations. In addition, it has been shown that different optimizers exhibit significantly different performance characteristics~\cite{Pellow-Jarman2021Comparison}. In this work, we consider four optimizers: ``COBYLA'', ``Powell'', ``BFGS'', and ``SLSQP''.

The paper is organized as follows. In Sec.~\ref{sec2}, we introduce the Hamiltonian to be studied using the VQE algorithm. Section~\ref{sec3} presents the details of our simulations, including the description of the ansatze, the optimizers employed, and the choice of initial guesses. In Sec.~\ref{sec4}, we compare the performance of the different algorithms for our specific problem across various system sizes. Both noiseless and noisy simulations are considered.

\section{Physical model}\label{sec2}

We study a one-dimensional system described by the following Hamiltonian
\begin{align}
H &= H_{\mathrm{kin}} + H_{sd} + H_{\mathrm{Coul}}, \label{eq:H_total} \\[4pt]
H_{\mathrm{kin}} &= -t \sum_{\langle i,i+1 \rangle, s} 
\left( c_{i,s}^\dagger c_{i+1,s} + \mathrm{h.c.} \right), \label{eq:H_kin} \\[4pt]
H_{sd} &= -J \sum_{i,s,s'} 
\left( \vec{B}_i \cdot c_{i,s'}^\dagger \, \vec{\sigma}_{s,s'} \, c_{i,s} \right), \label{eq:H_sd} \\[4pt]
H_{\mathrm{Coul}} &= U_c \sum_{i,\, s \neq s'} 
n_{i,s} \, n_{i,s'}, \label{eq:H_coul}
\end{align}
Here, $\vec{B}_i$ denotes the magnetization distribution of localized magnetic moments. These moments, originating from $d$-electrons, interact with free $s$-electrons via the $sd$ interaction. The constant $J$ represents the $sd$-exchange coupling, and $\vec{\sigma}_{s,s'}$ are the Pauli matrices. In addition, the free electrons experience an on-site Coulomb repulsion, characterized by the interaction constant $U_c$. The operators $c_{i,s}^\dagger$ and $c_{i,s}$ denote the creation and annihilation of an electron at site $i$ with spin $s$, while $n_{i,s}$ is the corresponding electrons number operator. The kinetic energy $E_{\text{kin}}$ is described by a tight-binding Hamiltonian with nearest-neighbor hopping constant $t$; only hopping between nearest neighbors is included. The system contains $N_s$ sites and a fixed number of electrons $N_e$. Open boundary conditions are considered.

We focus here on the ground state of the system. The number of degrees of freedom is determined by the number of sites multiplied by the spin degrees of freedom, i.e., $2N_s$. Modern quantum computers can handle up to about 100 qubits, which would suggest that systems with around 50 sites might be accessible. In practice, however, this is not the case due to limitations on circuit depth. The depth of the ansatz required to approximate the wave function grows rapidly with the number of qubits. The number of two-qubit gates (CZ or CX, depending on the hardware) largely determines the error in the final quantum state. Current quantum computers can reliably execute only several thousand two-qubit gates, which strongly restricts the size of the tight-binding models that can be simulated. In this work, we therefore study systems with fewer than 10 sites. Hamiltonians for such system can be readily diagonalized on classical computers. This can serve as ground truth for precision estimation.

\section{Quantum simulations}\label{sec3}
We are using VQE algorithm to find the ground state of the Hamiltonian Eq.~(\ref{eq:H_total}). This hybrid algorithm assumes that a quantum state $\Psi$ is prepared by the quantum circuit using some parametric ansatz. The ansatz is described by the parameters set \{$\theta_k$\}. Then the energy of this quantum state is calculated $E_{\Psi} = \langle \Psi | H | \Psi \rangle$. Multiple measurements are performed after the preparation of the quantum state to estimate different terms of the system Hamiltonian to finally get the energy value. This procedure is repeated multiple time for various parameters set \{$\theta_k$\}. Minimizing the energy over the parameters set one finds the system ground state. 

In this work, we employ several different ansatze to construct the quantum state and apply various optimization procedures to obtain the ground state of the system.

The Hamiltonian in Eq.~(\ref{eq:H_total}) is initially expressed in terms of electronic creation and annihilation operators. To implement it on a quantum computer, it must be mapped onto qubit operators. Several mappings are available for this purpose~\cite{Bravyi2002FermionicQC, Seeley2012JWTransform, PhysRevLett.86.1082, Verstraete2005MPS}. In this study, we adopt the Jordan-Wigner transformation~\cite{PhysRevLett.86.1082}.

\subsection{Correspondence between Hamiltonian states and qubits}\label{subsec_3_1}

In this problem, an electron can occupy any of the $2N_s$ states (two spin states per site). Within the Jordan-Wigner transformation~\cite{PhysRevLett.86.1082, Tranter2018Comparison}, each of these states is mapped to a qubit in a quantum computer. We assume that spin-up states ($s=1$) are represented by qubits $(0, \ldots, N_s - 1)$, while spin-down states are represented by qubits $(N_s, \ldots, 2N_s - 1)$. A physical state $(i,s)$ is considered occupied if the corresponding qubit is in the state $\lvert 1 \rangle$.

\subsection{Ansatze}\label{subsec_3_2}

We tested several different ansatze to determine which performs best. Each ansatz is characterized by the number of parameters it requires and by its circuit depth, which we quantify through the number of CX gates. The performance of an ansatz is assessed in terms of the number of iterations needed for the variational procedure to converge to the target ground-state energy $E_{g}^{qc}$, the achieved energy precision, and the fidelity of the resulting ground-state wave function. The ground-state energy obtained by direct diagonalization on a classical computer, $E_{g}^{dd}$, serves as the reference.

\subsubsection{Generic ansatz}\label{subsec_3_2_1}

\begin{figure}[h]
\centering
\includegraphics[width=1\textwidth]{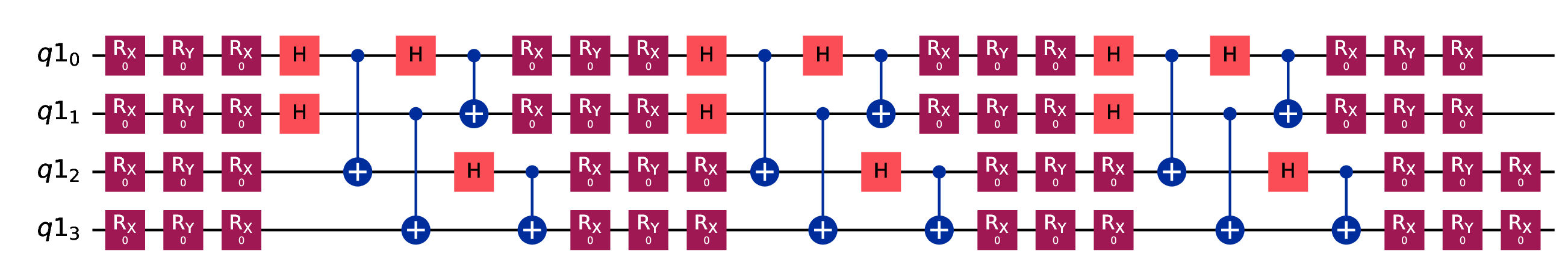}
\caption{An example of a generic ansatz for a four-qubit circuit is shown. Three repeating layers are illustrated. This four-qubit circuit corresponds to a system with two sites and is used to construct the associated quantum ``wave function''.}\label{fig_generic_ansatz}
\end{figure}

The generic ansatz is constructed from repeating blocks of rotation and entangling gates. A typical example of such an ansatz is shown in Fig.~\ref{fig_generic_ansatz}. The circuit illustrates the four-qubit case corresponding to a system with two sites: the top two qubits represent spin-up states, while the bottom two correspond to spin-down states. Tree blocks of rotations and entanglement are used. The figure, generated using the Qiskit quantum programming framework, employs standard quantum circuit notations. Here, $R_x$ denotes a qubit rotation around the $x$-axis by an angle $\theta$, $R_y$ represents a rotation around the $y$-axis, and $H$ is the Hadamard gate. The two-qubit operation marked with a white cross corresponds to the CNOT gate. The ansatz is parameterized by the set of rotation angles ${\theta_k}$ assigned to each rotation gate. In the example of Fig.~\ref{fig_generic_ansatz}, 48 rotation gates are present, requiring 48 parameters to fully describe the ansatz. 

A drawback of this generic ansatz is that it does not conserve the total number of particles in the system. To address this issue, we introduce an additional term in the Hamiltonian that enforces conservation of the electron number:

\begin{equation}
H_{\text{pc}}=E_\text{f}\left(n_{\text{tot}}-n_e\right)^2, \label{Eq_particle_cons_term}
\end{equation}
where, $E_\text{f}$ denotes the energy penalty for adding or removing a particle from the system, $n_e$ is the target (expected) number of electrons, and $n_{\text{tot}}$ is the operator corresponding to the total number of electrons in the system.

\subsubsection{Cluster ansatz}\label{subsec_3_2_2}

The cluster ansatz is well described in Refs.~\cite{Romero2018Strategies, PhysRevA.98.022322}. This ansatz conserves the total number of particles and is built from single-particle transitions of the form $c_{i,s}^\dagger c_{i^\prime,s^\prime}$ and double-particle transitions of the form $c_{i,s_1}^\dagger c_{j,s_2}^\dagger c_{k,s_3} c_{l,s_4}$. The fermionic operators are mapped to qubit operators via the Jordan-Wigner transformation. Evolution operators are then constructed based on these transition operators, allowing for partial transitions between states. The ``amount'' of the wave function transferred from one state to another is controlled by the rotation angles $\theta_i$, where $i$ indexes all possible single- and double-particle transitions in the system. By varying the set of angles ${\theta_i}$, one can generate arbitrary superpositions of Slater determinants.

\subsubsection{Modifies cluster ansatz (YAB)}\label{subsec_3_2_3}

Yordanov et al.~\cite{Yordanov2020EfficientCircuits} proposed a modified cluster ansatz that uses fewer CX gates and allows for a shallower quantum circuit. We will call this ansatz YAB ansatz using first characters of the last names of the paper authors. In general, the YAB ansatz can reduce the circuit depth by up to a factor of three compared to the standard cluster ansatz.

\begin{figure}[ht]
    \centering
    \begin{subfigure}{1\textwidth}
    	\caption{Single electron transition. Original YAB ansatz.}
        \includegraphics[width=\linewidth]{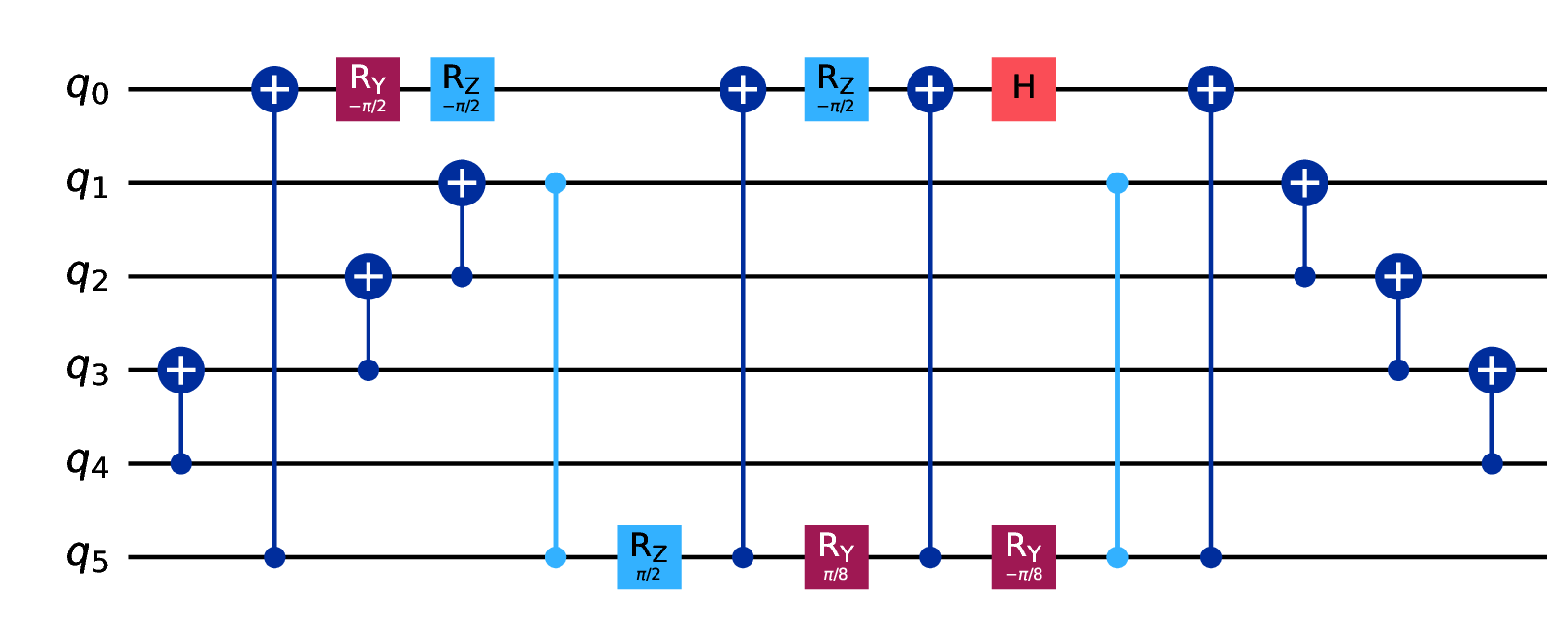}
        
    \end{subfigure}
    \hfill
    \begin{subfigure}{0.6\textwidth}
    	\caption{Single electron transition. Simplified YAB ansatz.}
        \includegraphics[width=\linewidth]{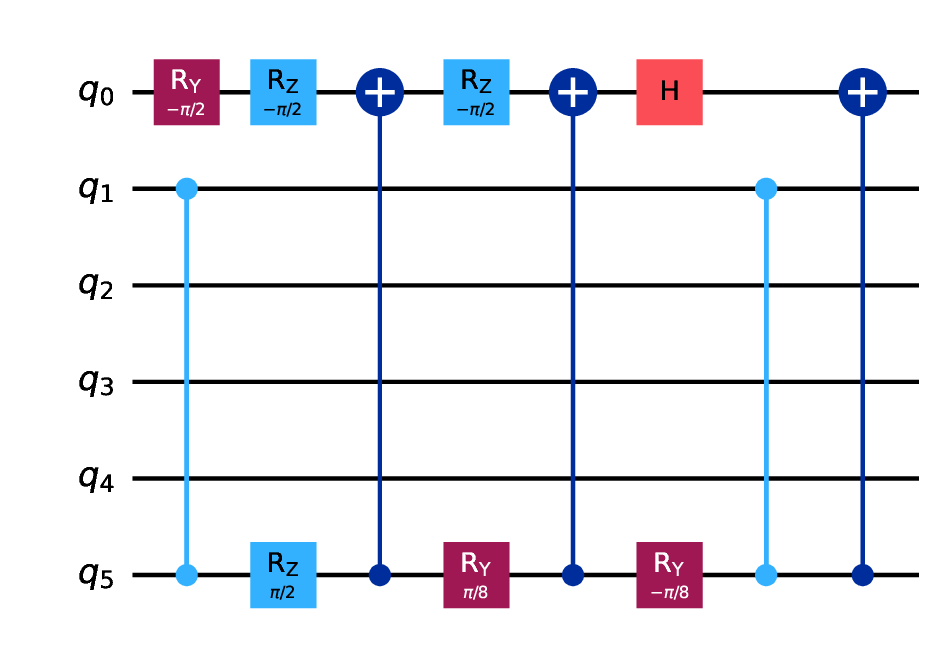}
        
    \end{subfigure}
    \caption{(a) Single-electron transition implemented using the circuit proposed in Ref.~\cite{Yordanov2020EfficientCircuits}. (b) Modified circuit design that omits the CX-gate ladder.}\label{fig_simplified_Yordanov}
\end{figure}

\subsubsection{Simplified YAB ansatz}\label{subsec_3_2_4}

The Jordan-Wigner transformation maps fermionic operators to qubit operators in a way that preserves parity and ensures the correct phase (sign) of the wave function after a transition. However, in our ansatz, the rotation angles $\theta_i$ control the phase of the final state after each transition. Therefore, strictly preserving the sign within the transition operator is not necessary: by adjusting $\theta_i$, we can still generate all possible system states even if the transition operator does not reproduce the exact fermionic sign.

To maintain fermionic parity, the original transition block includes a staircase of CX-gates. In our modified ansatz, we remove this CX-gate ladder and test its effect. This modification substantially reduces the number of CX-gates and results in a shallower quantum circuit. Fig.~\ref{fig_simplified_Yordanov} illustrates a single-electron transition block from the original YAB ansatz and our modified version without the CX-gate ladder. The example shows the transition between states 0 and 5. The number of CX-gates in the single-electron transition is significantly reduced; the farther apart the states, the more CX-gates can be removed. For neighboring states, no reduction is possible. A similar CX-gate ladder removal can be applied to double-electron transitions.

\subsubsection{Single electron transitions Vs Double electron transitions}\label{subsec_3_2_5}
Consider an initial system wave function with a given number of occupied states. The ansatz transforms this state into another one. Typically, this can be achieved using either single-electron transitions alone or a combination of single- and double-electron transitions. On one hand, including double-electron transitions allows the desired state to be reached with fewer overall transitions; on the other hand, each double transition requires more quantum gates. Here, we compare these two cases. In the first case, only single-electron transitions are used. The ansatz consists of a sequence of single transitions applied in the following order: $0 \rightarrow 1$, $0 \rightarrow 2$, ..., $1 \rightarrow 2$, $1 \rightarrow 3$, and so on. Each type of transition appears exactly once and at a specific position in the sequence. In the second case, double-electron transitions are also included. Again, both single- and double-electron transitions of each type appear only once within the ansatz.

\subsubsection{Comparison of various ansatze}\label{subsec_3_2_6}
Below we compare different ansatze in terms of the number of CX gates. The systems considered contain 2, 3, or 4 sites. It can be seen that for the cluster ansatz including both single and double transitions, the number of CX gates grows rapidly with system size and becomes impractical for current NISQ devices even at four sites. Using the YAB ansatz reduces the number of CX gates by approximately a factor of three. The simplified YAB ansatz proposed in this work achieves an additional reduction of about 20-25\% in CX gate count.

Interestingly, restricting the ansatz to single-electron transitions only leads to a drastic reduction in the number of gates - by roughly an order of magnitude. The key question is whether an ansatz based solely on single transitions can still span a sufficiently large portion of the solution space. Repeating the single-transition ansatz several times (e.g., three repetitions) expands the accessible state space while keeping the CX gate count significantly lower than that of any ansatz including double transitions. In what follows, we investigate whether the accuracy of the single-transition ansatz can approach that of the double-transition version.

\begin{table}[htbp]
\caption{Comparison of circuit depth for various anzatse. ``S'' denotes the ansatz with single-electron transitions only, while ``D'' denotes the ansatz including double transitions. ``3S'' corresponds to the single-transition ansatz repeated three times. $N_s$ is the number of nodes, $N_p$ is the number of variational parameters in the ansatz, $N_{CX}$ is the number of CX-gates.}\label{tab2}
\begin{tabular}{|l|c|c|c|c|c|c|}
\hline
Ansatz type & $N_s$ & $N_p$ & $N_{CX}$ \\
\hline
Generic  & 2 & 48 & 16 \\
Cluster, SD  & 2 & 9  & 184\\
YAB, SD  & 2 & 9  & 86\\
YAB, S  & 2 & 9  & 34\\
Simplified YAB, SD  & 2 & 9  & 76\\
Simplified YAB, S  & 2 & 6  & 20\\
Simplified YAB, 3S  & 2 & 18  & 60\\
\hline
Generic  & 3 & 72 & 21 \\
Cluster, SD  & 3 & 60  & 2876\\
YAB, SD  & 3 & 60  & 983\\
YAB, S  & 3 & 15  & 107\\
Simplified YAB, SD  & 3 & 60  & 812\\
Simplified YAB, S  & 3 & 15  & 57\\
Simplified YAB, 3S  & 3 & 45  & 171\\
\hline
Generic  & 4 & 96 & 30 \\
Cluster, SD  & 4 & 238  & 15792\\
YAB, SD  & 4 & 238  & 4776\\
YAB, S  & 4 & 28  & 210\\
Simplified YAB, SD  & 4 & 238  & 3572\\
Simplified YAB, S  & 4 & 28  & 114\\
Simplified YAB, 3S  & 4 & 28  & 342\\

\hline
\end{tabular}
\end{table}

\subsection{Choice of the initial guess}\label{subsec_3_3}
The convergence of the optimization procedure can strongly depend on the choice of the initial guess. In this work, we consider a one-dimensional chain of sites hosting localized magnetic moments and itinerant electrons. The nature of the ground state varies significantly with the system parameters. For weak exchange interaction, an antiferromagnetic (AFM) ordering of spins is expected, whereas strong s-d coupling favors a ferromagnetic (FM) ground state. The on-site Coulomb repulsion also plays a crucial role: when it is strong, electrons tend to remain localized on different sites, while weak Coulomb interaction allows them to delocalize across the entire chain.

\begin{figure}[htbp]
    \centering
    \begin{subfigure}{0.17\textwidth}
        \centering
        \includegraphics[width=\linewidth]{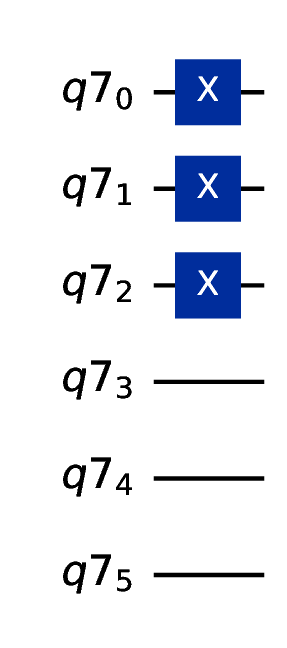}
        \caption{FM state}
        \label{fig:init1}
    \end{subfigure}
    \hfill
    \begin{subfigure}{0.5\textwidth}
        \centering
        \includegraphics[width=\linewidth]{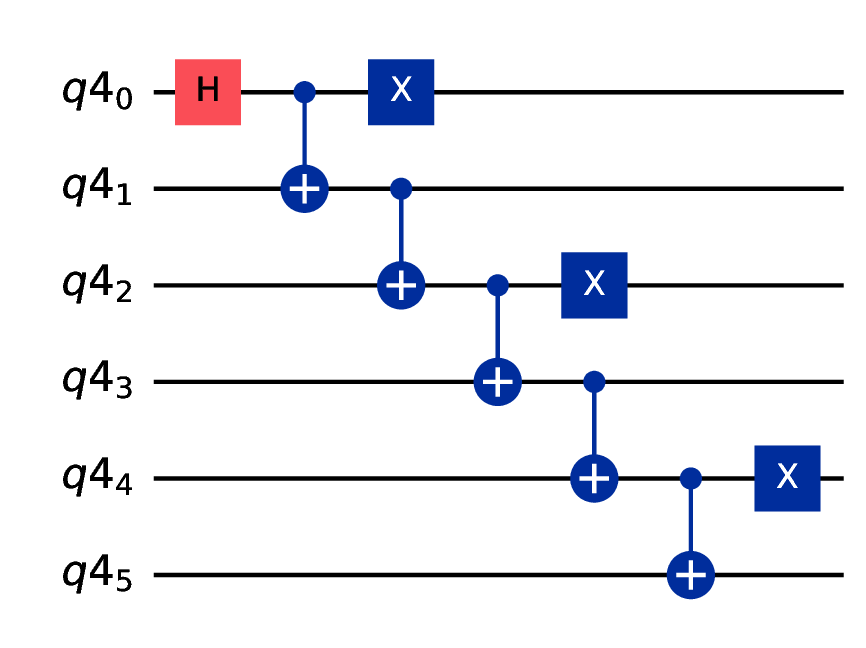}
        \caption{Double AFM state}
        \label{fig:init2}
    \end{subfigure}
    \hfill
    \begin{subfigure}{0.17\textwidth}
        \centering
        \includegraphics[width=\linewidth]{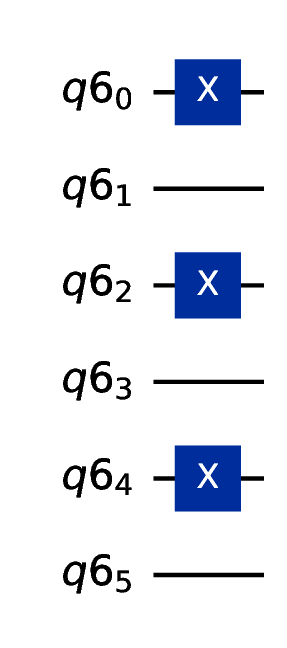}
        \caption{AFM state}
        \label{fig:init3}
    \end{subfigure}
    \caption{The circuits creating various initial guesses.}
    \label{fig_init_states}
\end{figure}

To accelerate convergence and avoid trapping in incorrect local minima, we employ different types of initial guess states depending on the parameter regime. An additional circuit is appended at the beginning of the variational algorithm to prepare the chosen initial state. Figure~\ref{fig_init_states} illustrates three examples of initial-state preparation circuits used in this work for a system with three sites and three electrons (half-filling). The left panel corresponds to a simple FM configuration, where all three electrons occupy spin-up states at sites 1, 2, and 3. Such an initial guess is suitable for strong s-d coupling, where the localized spins are uniformly magnetized. The right panel depicts an AFM configuration with one electron per site, but with the middle electron occupying a spin-down state. This initial guess is appropriate when the kinetic energy and Coulomb repulsion dominate, but an external field slightly favors one of the AFM orientations. Finally, the middle panel shows a superposition of two opposite AFM configurations, which corresponds to the case where the s-d interaction is negligible.

\section{Ansatze performance comparison}\label{sec4}
\subsection{Case of 2 nodes. Noiseless simulation}\label{subsec_4_1}
To compare the performance of different ansatze, we employ a test Hamiltonian for a two-site system. Three sets of parameters were considered:  
1) $U_c/t = 10$ ($t = 1$), $J = 0.2\,t$;  
2) $U_c/t = 2.5$ ($t = 4$), $J = 0.025\,t$;  
3) $U_c/t = 0.1$ ($t = 1$), $J = 0.2\,t$.  

We study the half-filling case with two electrons. The parameters are chosen such that the ground state is non-degenerate. The corresponding ground-state energies are  
$E_{g1} = -10.4\,t$, $E_{g2} = -3.2\,t$, and $E_{g3} = -2.002\,t$.  

\begin{table}[htbp]
\centering
\caption{Performance of various ansatze. Task with 2 nodes. Averaging is performed over 9 set of parameters. No noise. $N_{it}$ is the iteration number to get the result. $T_{calc}$ is the simulation time. $dE$ is the relative deviation of the energy obtained with VQE against the direct diagonalization. $F$ is the wave function fidelity. }
\label{tab:A1}
\begin{tabular}{|l|c|c|c|c|c|}
\hline
Ansatz & Optimizer & $N_{it}$ & $T_{calc}$ (s) & $dE$ (\%) & $F$  \\
\hline
Cluster, SD           & COBYLA  & 574 & 15      & 0.02  & 0.994  \\
Cluster, SD          & Powell  & 532  & 14   & 0.005  & 0.9993 \\
Cluster, SD           & BFGS  & 255  & 6    & 0.21 & 0.8889 \\
Cluster, SD           & SLSQP  & 161  & 4    & 0.0 & 1 \\
\hline
Generic                        & COBYLA & 45740   & 309    & 0.06 & 0.98 \\
Generic                        & Powell & 10900	  & 72    &  0.35 & 0.944 \\
Generic                        & BFGS & 10130   & 65 &  0.0 & 0.9999 \\
Generic                        & SLSQP & 6640 & 43 & 0.002 & 0.9999 \\
\hline
YAB, SD & COBYLA  & 860  & 10   & 0.009  & 0.9995 \\
YAB, SD & Powell  & 657  & 8   & 0.23 & 0.899 \\
YAB, SD & BFGS  & 316  & 3.8    & 0.0 & 1 \\
YAB, SD & SLSQP  & 212  & 2.5    & 0.0 & 1 \\
\hline
Simplified YAB, SD & COBYLA  & 683  & 8.1   & 0.03  & 0.993 \\
Simplified YAB, SD & Powell  & 529  & 6.2   & 0.22 & 0.892 \\
Simplified YAB, SD & BFGS  & 225  & 2.6    & 0.0 & 1 \\
Simplified YAB, SD & SLSQP  & 176  & 2.0    & 0.0 & 1 \\
\hline
Simplified YAB, S     & COBYLA  & 628  & 4.2      & 0.2 & 0.92 \\
Simplified YAB, S     & Powell  & 405  & 2.6    & 0.2 & 0.89 \\
Simplified YAB', S     & BFGS  & 248  & 1.6   & 0.14 & 0.96 \\
Simplified YAB, S     & SLSQP   & 144  & 0.9    & 0.14 & 0.96  \\
\hline
Simplified YAB, 3S     & COBYLA  & 770  & 11.7      & 0.22 & 0.89 \\
Simplified YAB, 3S      & Powell  & 740  & 11.1    & 0.22 &  0.9\\
Simplified YAB, 3S      & BFGS  & 490  & 7    & 0 & 1 \\
Simplified YAB, 3S      & SLSQP   & 320  & 3.5    & 0 & 1 \\
\hline
\end{tabular}
\end{table}

The wavefunctions of these three states are qualitatively different.  
1) For the first parameter set, the ground state is ferromagnetic (FM): $\uparrow\uparrow$.  
2) For the second set, the ground state is primarily a combination of two antiferromagnetic (AFM) states: $ 0.67(\uparrow\downarrow + \downarrow\uparrow) + 0.23(\updownarrow\_\_ + \_\_\updownarrow)$, where the first term corresponds to electrons occupying different sites, while the second represents both electrons residing on the same site. The sign $\updownarrow$ represents two electrons on the same node, and the symbol $\_\_$ indicates the empty site.
3) For the third set, the state is $ 0.51(\uparrow\downarrow + \downarrow\uparrow) + 0.49(\updownarrow\_\_ + \_\_\updownarrow)$.

We use three types of initial guesses for the variational optimization:  
1) a symmetric superposition of AFM states $(\uparrow\downarrow + \downarrow\uparrow)$;  
2) a single AFM configuration $\uparrow\downarrow$; and  
3) a FM configuration $\uparrow\uparrow$.

The results are averaged over the three Hamiltonians and three initial states, resulting in nine optimization tasks in total. The table below summarizes how various ansatze perform in combination with different optimization algorithms. We examine 6 ansatze and four optimizers (both gradient-based and gradient-free): COBYLA, Powell, BFGS, and SLSQP~\cite{Pellow-Jarman2021Comparison}. Ideal, noise-free simulations were used. The table reports the average relative deviation of the final energy from the exact ground-state value $dE$, as well as the number of iterations, $N_{it}$ (and total computation time, $T_{calc}$) required to achieve convergence. The fidelity of the obtained wave function $F$ is also shown in the table.

Based on the presented data, we can conclude that the BFGS and SLSQP optimizers perform, on average, better than COBYLA and Powell. Almost all ansatze yield high fidelity and accurate ground-state energy estimates when used with these optimizers. The results also show that the simplified cluster ansatze perform as well as, or even better than, the original cluster ansatz - except for the simplest single-transition version.

The simplified YAB ansatz, consisting of single transitions only and repeated three times, requires the fewest gates among all physics-inspired ansatze, yet achieves performance comparable to the full cluster ansatz, which uses approximately three times more CX gates.

The generic ansatz, when combined with the particle-number-conserving term in the Hamiltonian, also performs on par with the physics-inspired ansatze. In terms of the number of CX gates, it is the most efficient option for execution on real quantum hardware, as fewer entangling gates reduce accumulated errors. However, this ansatz typically requires a significantly larger number of optimization iterations and longer total computation time.

\begin{table}[htbp]
\centering
\caption{Performance of various ansatze. Task with 2 nodes. Averaging is performed over 9 set of parameters. No noise. $N_{it}$ is the iteration number to get the result. $T_{calc}$ is the simulation time. $dE$ is the relative deviation of the energy obtained with VQE against the direct diagonalization. $F$ is the wave function fidelity.  SLSQP optimizer is used}
\label{tab:A2}
\begin{tabular}{|l|c|c|c|c|c|}
\hline
Ansatz & $N_s$ & $N_{it}$ & $T_{calc}$ (s) & $dE$ (\%) & $F$  \\
\hline
Cluster, SD           & 3  & 4540  & 1430   & 0.0 & 1 \\
Cluster, SD           & 4  & 22500  & 35900   & 0.0 & 1 \\
\hline
Generic                        & 3 & 15500 & 160 & 6 & 0.6 \\
Generic                        & 4 & 29100 & 400 & 17 & 0.4 \\
\hline
YAB, SD & 3  & 6000  & 600    & 0.0 & 1 \\
YAB, SD & 4  & 41100  & 31150    & 0.0 & 1 \\
\hline
Simplified YAB, SD & 3  & 4800  & 432    & 0.0 & 1 \\
Simplified YAB, SD & 4  & 21650  & 8575    & 0.0 & 1 \\
\hline
Simplified YAB, S     & 3   & 830  & 11    & 1.2 & 0.64  \\
Simplified YAB, S     & 4   & 2400  & 58    & 1.3 & 0.83  \\
\hline
Simplified YAB, 3S      & 3   & 2460  & 84    & 0 & 1 \\
Simplified YAB, 3S      & 4   & 12200  & 814    & 0.0006 & 0.99993 \\
\hline
\end{tabular}
\end{table}

\subsection{Case of 3 and 4 nodes. Noiseless simulation}\label{subsec_4_2}
We performed a similar test for systems with 3 and 4 nodes, considering the half-filling case in both instances. The Hamiltonian parameters are the same as in the previous section. Since SLSQP performed best for smaller systems, we use this optimizer exclusively here. Table~\ref{tab:A2} compares the performance of various ansatze for the 3- and 4-node systems. It can be seen that the generic ansatz underperforms compared to the physics-inspired ansatze, with even the single-transition ansatz achieving higher accuracy. The single-transition ansatz repeated three times performs comparably to the full cluster ansatz while requiring significantly fewer CX gates and achieving computation times approximately 20-40 times faster.

\subsection{Estimate of IBM QC noise level}\label{subsec_4_3}
Noiseless simulations were used in the previous section. On a real quantum computer, however, noise can significantly affect the performance of an ansatz. Performing all of the above tests on actual quantum hardware is both time-consuming and costly. Therefore, we estimate the noise level to be added to simulations in order to approximate the behavior of a real quantum device. As an example, we used the IBM quantum computer \texttt{ibm-brisbane}. We compared the energy calculated on the IBM QC for a fixed wave function (fixed ansatz parameters) with noisy simulations of the same experiment. For classical simulations, we employed the Qiskit AER simulator with and without added noise. The noise model includes errors on CX gates only. Two noise levels were tested: 1\% and 2\%.  It should be noted that the performance of the real QC also depends on the transpilation process. The transpilation parameter known as the ``optimization level'' affects the final energy calculation. For the \texttt{ibm-brisbane} device and the algorithm considered here, the optimal value was found to be ``optimization level = 2''.

\begin{table}[h!]
\centering
\caption{Estimated energy (in units of $t$) for the state prepared using the simplified YAB ansatz. Energies are computed using different methods, including noiseless classical simulations, noisy classical simulations, and measurements on a real quantum computer.}
\label{tab:afm_energy}
\begin{tabular}{|c |c| c| c |c|}
\hline
\textbf{$N_s$} & \textbf{Noiseless simulations} & \textbf{Noise level, (\%)} & \textbf{Noisy simulations}  & \textbf{IBM QC} \\
\hline
1 & -4.9 & 2& -4.61  & -4.87   \\
2 & -10.0& 2& -7.39  & -8.57  \\
3 & -14.9 & 2 & -8.62& -11.38  \\
\hline
1 & -4.9 &1& -4.75 & -4.87  \\
2 & -10.0&1 & -8.60& -8.57  \\
3 & -14.9&1& -11.35 & -11.38  \\
\hline
\end{tabular}
\end{table}

To compare classical simulations with the real IBM QC, we use the shortest ansatz, namely the simplified YAB ansatz including both single and double transitions. In this test, all rotation angles $\theta_i$ are set to zero. This does not imply that no gates are applied; the full set of gates is executed as in a typical run with arbitrary $\theta_i$. The initial guess is the single AFM state. No variational optimization is performed; only a single energy estimate is obtained. Table~\ref{tab:afm_energy} presents a comparison of the three simulation methods.

From the table, we conclude that the \texttt{ibm-brisbane} QC can be reasonably approximated by a noise model including 1\% CX gate errors. It is evident that the noise level of the real device is significant. For the 3-node system with an ansatz containing approximately 800 CX gates, the energy estimation error is on the order of 30\% ($\sim 3.5\,t$), while the energy level spacing can be as small as $0.1\,t$.

\begin{table}[htbp]
\centering
\caption{Performance comparison of different ansatze for the 2-node system under 1\% CX gate noise. Results are averaged over 9 parameter sets. $N_{\rm it}$: number of iterations, $T_{\rm calc}$: simulation time, $dE$: relative energy deviation from exact diagonalization, $F$: wave function fidelity.
}
\label{tab:A3}
\begin{tabular}{|l|c|c|c|c|c|}
\hline
Ansatz & Optimizer & $N_{it}$ & $T_{calc}$ (s) & $dE$ (\%) & $F$  \\
\hline
Cluster, SD           & COBYLA  & 950 & 110      & 78  & 0.94  \\
Cluster, SD          & Powell  & 1350  & 162   & 77  & 0.95 \\
Cluster, SD           & BFGS  & 888  & 96    & 77 & 0.87 \\
Cluster, SD           & SLSQP  & 470  & 51    & 76 & 0.9995 \\
\hline
Generic                        & COBYLA & 29700   & 1330    & 27 & 0.667 \\
Generic                        & Powell & 20100	  & 658    &  28 & 0.67 \\
Generic                        & BFGS & 22000   & 695 &  26 & 0.67 \\
Generic                        & SLSQP & 10025 & 340 & 27 & 0.78 \\
\hline
YAB, SD & COBYLA  & 7360  & 455   & 40  & 0.9998 \\
YAB, SD & Powell  & 1080  & 64   & 40 & 0.999 \\
YAB, SD & BFGS  & 657  & 40    & 39 & 0.9997 \\
YAB, SD & SLSQP  & 560  & 33    & 40 & 0.9996 \\
\hline
Simplified YAB, SD & COBYLA  & 8130  & 525  & 36  & 0.9998 \\
Simplified YAB, SD & Powell  & 1350  & 77  & 36 & 0.989 \\
Simplified YAB, SD & BFGS  & 692  & 41    & 36 & 0.9999 \\
Simplified YAB, SD & SLSQP  & 422  & 23   & 37 & 0.9999 \\
\hline
Simplified YAB, S     & COBYLA  & 4540  & 165      & 13 & 0.889 \\
Simplified YAB, S     & Powell  & 300  & 10    & 14 & 0.889 \\
Simplified YAB, S     & BFGS  & 413  & 14   & 14 & 0.96 \\
Simplified YAB, S     & SLSQP   & 248  & 9    & 14 & 0.78  \\
\hline
Simplified YAB, 3S     & COBYLA  & 24505  & 1740      & 31 & 0.9994 \\
Simplified YAB, 3S      & Powell  & 4060  & 270    & 32 &  0.999\\
Simplified YAB, 3S      & BFGS  & 3500  & 230    & 30 & 0.9994 \\
Simplified YAB, 3S      & SLSQP   & 2050  & 130    & 30 & 0.998 \\
\hline
\end{tabular}
\end{table}

\subsection{Comparison of various ansatze in the presence of noise in the simulations}\label{subsec_4_4}
Here, we use noisy simulations to assess how the studied ansatze might perform on real NISQ-era quantum computers. Typically, the most significant errors arise from two-qubit gates, such as the CX gates in our case. To simulate this, we introduce noise only on the CX gates, specifically a depolarizing error with 1\% probability. We perform the same tests of the ansatze and optimization algorithms as described in Sec.~\ref{subsec_4_1}. The results are summarized in Table~\ref{tab:A3}.

Several interesting features emerge from the table. First, while the predicted ground-state energies exhibit large errors (ranging from 14\% to 70\%), the fidelities of the wave functions obtained using the VQE algorithm remain relatively high. The best wave function performance is observed for the YAB ansatze (except for the algorithm using only single transitions). For these ansatze, the ``COBYLA'' optimizer requires the largest number of iterations and the longest computation time. Interestingly, the single-transition ansatz repeated three times achieves the highest accuracy in terms of energy prediction, likely because it uses the fewest CX gates, minimizing gate-induced errors. This is further supported by the observation that the generic ansatz and the single-transition-only ansatz exhibit even lower energy deviations. Note that while fewer CX gates improve energy prediction, they reduce wave function fidelity. However, wave function accuracy is arguably more important: knowing the correct wave function allows one to compute the energy classically, even if the quantum computer cannot provide an accurate energy estimate, whereas knowing the energy does not allow reconstruction of the wave function.

Another observation is that, in contrast to noiseless simulations, all optimization algorithms achieve similar accuracy in energy and wave function fidelity under noisy conditions. Nevertheless, some optimizers require more time to converge; ``COBYLA'' is the slowest. Overall, optimization convergence is slower in the presence of noise, typically requiring more iterations than in the noiseless case.

\section{Conclusion}\label{sec7}
We use the VQE algorithm to study the ground state of a lattice model with on-site Coulomb repulsion, nearest-neighbor hopping, and on-site $sd$ interaction. We compared the performance of several ansatze, including cluster and generic ansatze, and tested multiple implementations of the cluster ansatz with varying circuit depths. Additionally, we proposed a modification of the cluster ansatz that reduces the number of two-qubit gates.  

We also compared different classical optimizers within the VQE algorithm, namely COBYLA, SLSQP, Powell, and BFGS. The performance of these algorithms was evaluated using both noiseless and noisy simulations. The noise level for the simulations was estimated based on measurements from the IBM \texttt{brisbane} quantum computer. For noiseless simulations, the best results were obtained with SLSQP and BFGS optimizers.  

The cluster ansatz was modified in two ways: (1) the CX-gate ladder in the transition operators was removed, and (2) the number of transition operators was optimized. We found that applying only single transition operators, but repeating them multiple times, significantly reduces the number of CX gates (from approximately 15,000 for the standard cluster ansatz with single and double transitions for an 8-orbital model to about 300 for the modified ansatz with single transitions repeated three times) while maintaining the same energy calculation precision.  

For noisy simulations, the shorter ansatz achieves better energy accuracy due to the reduced circuit depth. Interestingly, while the energy prediction precision decreases under noise, the fidelity of the wave function remains largely unaffected.

\backmatter

\section*{Declarations}

\begin{itemize}
\item Funding

Nothing to declare.

\item Conflict of interest/Competing interests 

On behalf of all authors, the corresponding author states that there is no conflict of interest.

%\item Ethics approval and consent to participate

%\item Consent for publication
\item Data availability

No datasets were generated or analyzed during the current study.

\item Code availability 

https://github.com/OlegUdalov/QC-qiskit-codes/tree/main/VQE

\end{itemize}

%%===================================================%%
%% For presentation purpose, we have included        %%
%% \bigskip command. Please ignore this.             %%
%%===================================================%%

%%=============================================%%
%% For submissions to Nature Portfolio Journals %%
%% please use the heading ``Extended Data''.   %%
%%=============================================%%

%%=============================================================%%
%% Sample for another appendix section			       %%
%%=============================================================%%

%% \section{Example of another appendix section}\label{secA2}%
%% Appendices may be used for helpful, supporting or essential material that would otherwise 
%% clutter, break up or be distracting to the text. Appendices can consist of sections, figures, 
%% tables and equations etc.

%\end{appendices}

%%===========================================================================================%%
%% If you are submitting to one of the Nature Portfolio journals, using the eJP submission   %%
%% system, please include the references within the manuscript file itself. You may do this  %%
%% by copying the reference list from your .bbl file, paste it into the main manuscript .tex %%
%% file, and delete the associated \verb+\bibliography+ commands.                            %%
%%===========================================================================================%%

\bibliography{VQE}% common bib file
%% if required, the content of .bbl file can be included here once bbl is generated
%%\input sn-article.bbl

\end{document}